\documentclass[amsmath,amssymb,twocolumn,showpacs]{revtex4}

\usepackage{graphicx}
\usepackage[usenames]{color}

\begin{document}

\title{Chemical pressure and hidden one-dimensional behavior
in rare earth tri-telluride charge density wave compounds}
\author{A. Sacchetti and L. Degiorgi} \affiliation{Laboratorium f\"ur
Festk\"orperphysik, ETH - Z\"urich, CH-8093 Z\"urich,
Switzerland.}
\author{T. Giamarchi} \affiliation{DPMC, University of Geneva, 24,
quai Ernest-Ansermet, CH1211 Geneva 4, Switzerland.}
\author{N. Ru and I.R. Fisher}
\affiliation{Geballe Laboratory for Advanced Materials and
Department of Applied Physics, Stanford University, Stanford,
California 94305-4045, USA.}

\date{\today}

\begin{abstract}
We report on the first optical measurements of the rare-earth
tri-telluride charge-density-wave systems. Our data, collected
over an extremely broad spectral range, allow us to observe both
the Drude component and the single-particle peak, ascribed to the
contributions due to the free charge carriers and to the
charge-density-wave gap excitation, respectively. The data
analysis displays a diminishing impact of the charge-density-wave
condensate on the electronic properties with decreasing lattice
constant across the rare-earth series. We propose a possible
mechanism describing this behavior and we suggest the presence of
a one-dimensional character in these two-dimensional compounds. We
also envisage that interactions and umklapp processes might play a
relevant role in the formation of the charge-density-wave state in
these compounds.
\end{abstract}


%\keywords{rare-earth,tellurides}

\maketitle

\section{Introduction}

Low dimensionality is an important issue in solid state physics
\cite{LeoBook}, owing to the general tendency of low-dimensional
systems to form charge- and spin-density-wave (CDW and SDW) states
\cite{CDW,peierls}. CDW and SDW phases are broken symmetry
ground-states driven by the electron-phonon and electron-electron
interactions, respectively. These phases, from the electronic
point of view, are induced by nesting of the Fermi surface (FS)
\cite{CDW}. Besides the single particle gap excitation, the
density-wave states are then characterized by the formation of the
collective CDW or SDW condensate \cite{CDW}. Density-waves have
been observed in several materials such as linear-chain organic
and inorganic compounds \cite{CDW}. Strong interest in
low-dimensional systems has also been brought about by the
considerable deviations of their normal state properties from
those of a Fermi liquid \cite{Jerome,Voit,Lutt,LeoEPJB}. In one
dimension, the Tomonaga-Luttinger-liquid or Luther-Emery
scenarios, implying phenomena like spin-charge separation and
non-universal power-law behavior of the spectral functions, are
most suitable \cite{LeoEPJB}. In several cases, characteristic and
peculiar power-law behaviors were indeed observed in the
spectroscopic (optical and photoemission) properties of various
quasi one-dimensional materials \cite{Voit,Lutt,LeoEPJB}. It is
worth noting that, although the theory of density waves is well
established in the one dimensional (1D) case \cite{CDW}, little is
known about the two dimensional (2D) case.

Another class of compounds, which recently gained importance in
the study of density waves, are the rare earth tri-tellurides
$R$Te$_3$ ($R$ = La - Tm, excepting Eu \cite{Dimasi2}). These
systems exhibit an incommensurate CDW, stable across the available
rare earth series \cite{Dimasi}. All these compounds have the same
average crystal structure (belonging to the $Cmcm$ space group
\cite{struct}) made up of square planar Te sheets \cite{footnote}
and insulating corrugated $R$Te layers which act as charge
reservoirs for the Te planes. The lattice constant decreases on
going from $R=$ La to $R=$ Tm \cite{LattConst}, i.e. by decreasing
the ionic radius of the rare earth atom. Therefore, the study of
these compounds allows to investigate the CDW state as a function
of the unit-cell volume and in particular of the in-plane lattice
constant $a$, which is directly related to the Te-Te distance in
the Te-layers.

Metallic conduction occurs along the Te-sheets leading to highly
anisotropic transport properties. For instance, the ratio between
the in and out of plane conductivity can be as high as 3000
\cite{Dimasi2}. The rare-earth atom is trivalent for all members
of the series \cite{Dimasi2} and thus the electronic structure is
quite similar in all of them. Band structure calculations
\cite{Dimasi,Band,Band2} reveal that the electronic bands at the
Fermi level derive from the Te $p_x$ and $p_y$ in-plane orbitals,
leading to a very simple FS, large part of which is nested by a
single wave-vector $\vec{q}=(0,x),$ $x \simeq 0.29a^{*}$
($a^{*}=2\pi/a$) in the base-plane of the reciprocal lattice
\cite{Band2}. This nesting appears to be the driving mechanism for
the CDW instability \cite{ARPES2,ARPES3}.

The compounds are characterized by an unusually large CDW gap,
which was observed by angle resolved photoemission spectroscopy
(ARPES) measurements. The CDW gap ranges between 200 and 400 meV,
depending on the rare earth
\cite{ARPES2,ARPES3,ARPES1,ARPES4,XPSgap}. Consistent with these
large gap values, $R$Te$_3$ compounds are well within the CDW
state already at room temperature and the CDW transition
temperature ($T_{CDW}$) is believed to be even higher than the
melting point \cite{ARPES1}. Owing to the 2D character of these
compounds, the gap is not isotropic and shows a wave-vector
dependence \cite{ARPES1,ARPES3}. In particular, since the vector
$\vec{q}$ does not nest the whole FS, there are parts of it which
are not gapped. Therefore, the CDW state coexists with the
metallic phase due to the free charge-carriers in the ungapped
regions of the FS. The study of these compounds could give in
principle an important insight into the interplay between the
metallic state and the broken-symmetry CDW phase. Furthermore, the
presence of a single nesting vector defines a preferred
crystallographic direction for the development of the CDW state,
and leads to features typical of a 1D system, despite the 2D
character of these compounds.

Besides to the prototype 1D systems \cite{CDW}, optical
spectroscopy was already successfully applied to the study of 2D
chalcogenides such as NbSe$_2$ and TaSe$_2$, where anomalous
behavior of the carriers' scattering rate was observed in the CDW
phase \cite{Vescoli,Dordevic}, as well as NbSe$_3$ and TaSe$_3$,
where a polaronic scenario for the CDW was proposed
\cite{Perucchi1,Perucchi2}. Here, we describe the first
comprehensive optical study on $R$Te$_3$. Optical spectroscopy is
in general an ideal tool to study CDW systems \cite{CDW}, since it
is able to reveal the opening of the CDW gap in the charge
excitation spectrum. The optical signature of the CDW phase is in
fact a finite-frequency peak, ascribed to the transition from the
CDW condensate to a single particle (SP) state \cite{SPpeak}.
Moreover, because of the imperfect nesting, as typical for quasi
2D systems like $R$Te$_3$, the optical technique also reveals the
free-carriers contribution in terms of a Drude peak.

\section{Experiment and Results}

We report on optical reflectivity measurements carried out on
$R$Te$_3$ single crystals for representative members across the
rare earth series $R$ = La, Ce, Nd, Sm, Gd, Tb and Dy. Single
crystal samples were grown by slow cooling a binary melt, as
described elsewhere \cite{Ru}. Plate-like crystals up to several
mm in diameter were removed from the melt by decanting in a
centrifuge. The crystals could be readily cleaved between Te
layers to reveal clean surfaces for the reflectivity measurements.
Exploiting several spectrometers and interferometers, the optical
reflectivity $R(\omega)$ was measured for all samples from the
far-infrared (6 meV) up to the ultraviolet (12 eV) spectral range,
with light polarized parallel to the Te-planes. Details pertaining
to the experiments can be found elsewhere \cite{Wooten,dressel}.
\begin{figure}[!tb]
\center
\includegraphics[width=8.5cm]{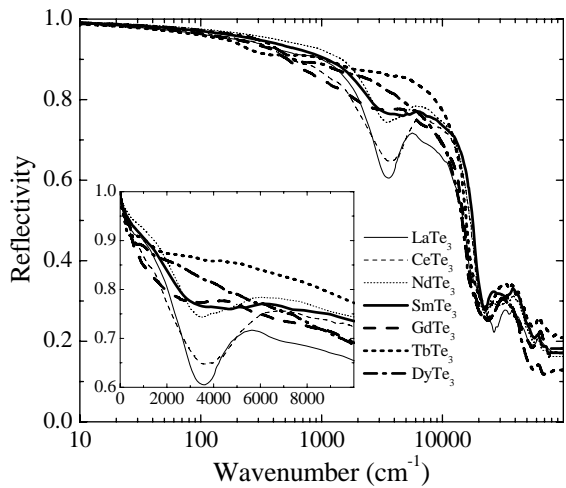}

\caption{$R(\omega)$ at room temperature of $R$Te$_3$ ($R$=La, Ce,
Nd, Sm, Gd, Tb and Dy). The inset shows a blow up of the
$R(\omega)$ in the energy range between 0 and 10000 cm$^{-1}$.}
\label{Refl}
\end{figure}

Figure 1 displays the overall $R(\omega)$ spectra for selected
members across the rare earth series. Consistently with the large
gap values, no temperature dependence of the spectrum was observed
between 2 K and 300 K. As expected from the presence of ungapped
regions of the FS, all samples exhibit a metallic $R(\omega)$,
tending to total reflection at zero frequency (i.e., $R(\omega)\to
1$, for $\omega\to 0$), and the appearance of a plasma edge around
10000~cm$^{-1}$. Above the plasma edge, several spectral features
are also observed. At lower frequency a bump is apparent in
$R(\omega)$ of all samples (see inset of Fig.~\ref{Refl}). This
feature is more evident in LaTe$_3$ but it becomes progressively
less pronounced and shifts to lower frequency on going from
LaTe$_3$ to DyTe$_3$, i.e., on decreasing the in-plane lattice
constant $a$.

The large explored spectral range allows us to perform reliable
Kramers-Kronig (KK) transformations. To this end, $R(\omega)$ was
extended towards zero frequency (i.e., $\omega\rightarrow 0$) with
the Hagen-Rubens extrapolation
($R(\omega$)=$1-2\sqrt{\omega/\sigma_{dc}}$) and with standard
power-law extrapolations at high frequencies \cite{Wooten}. The DC
conductivity values employed in the Hagen-Rubens extension of
$R(\omega)$ are consistent with transport measurements
\cite{Ru,Transp}. The KK transformations allow us to extract the
real part $\sigma_1(\omega)$ of the optical conductivity,
displayed in Fig. 2.

\section{Discussion}

\begin{figure}[!tb]
\center
\includegraphics[width=8.5cm]{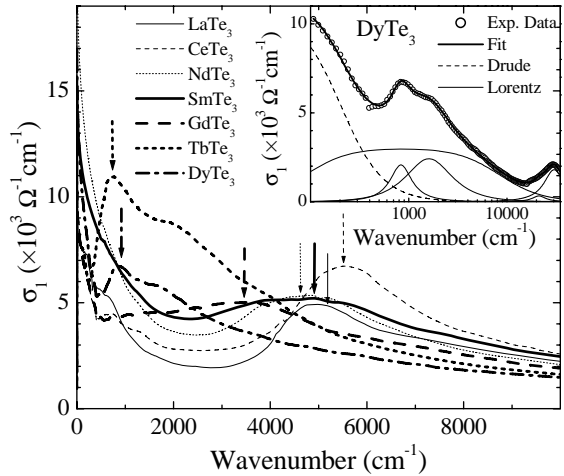}
\caption{Room temperature $\sigma_1(\omega)$ of $R$Te$_3$ ($R$=La,
Ce, Nd, Sm, Gd, Tb and Dy). Arrows mark the position of the SP
peak for each compound (see text). Inset: Drude-Lorentz fit for
DyTe$_3$, showing the experimental data, the fitted curve, the
Drude and the Lorentz components.} \label{Sig1}
\end{figure}

All $\sigma_1(\omega)$ spectra in Fig. 2 are characterized by the
presence of two main features, namely a Drude peak, revealing
metallic conduction due to the free charge-carriers, and a second
mid-infrared peak centered at finite frequency (arrows in
Fig.~\ref{Sig1}), corresponding to the bump observed in
$R(\omega)$ (Fig. 1). The depletion in the $\sigma_1(\omega)$
spectrum between these two features will be later identified with
the CDW gap. Higher energy excitations, corresponding to the
spectral features observed in $R(\omega)$ above the plasma edge,
are also present in $\sigma_1$ (not shown in the figure) and they
are ascribed to electronic interband transitions. The frequencies
of these excitations are compatible with the predictions from
band-structure calculations \cite{Band,Band2}, taking into account
the band energies at the $\Gamma$-point. On the other hand, these
band-structure calculations do not reveal any electronic
transition below 1 eV for the undistorted structure (i.e., in the
normal state). It is thus natural to identify the depletion in the
$\sigma_1(\omega)$ spectrum between the Drude and the mid-infrared
peaks with the CDW gap. Therefore, the mid-infrared peak is
ascribed to the charge excitation across the CDW gap into a single
particle (SP) state. In the following we will refer to this peak
as the SP peak.

Interestingly, our data show a clear red-shift of the SP peak from
LaTe$_3$ to DyTe$_3$. This effect can be observed directly from
the $\sigma_1(\omega)$ spectra (arrows in Fig.~\ref{Sig1}) or, to
a smaller extent, from $R(\omega)$ (inset of Fig.~\ref{Refl}). In
order to get more quantitative information, a fit procedure,
exploiting the phenomenological Drude-Lorentz model, was carried
out on all spectra. It consists in reproducing the dielectric
function by the following expression: {\setlength\arraycolsep{2pt}
\begin{eqnarray}
\nonumber \tilde{\epsilon}(\omega) & = & \epsilon_1(\omega)
+i\epsilon_2(\omega) =
\\ & = & \epsilon_{\infty}-\frac{\omega_P^2}{\omega^2+i \omega
\gamma_D}+\sum_j \frac{S_j^2}{\omega^2-\omega_j^2-i \omega
\gamma_j}
\end{eqnarray}}
where $\epsilon_{\infty}$ is the optical dielectric constant,
$\omega_P$ and $\gamma_D$ are the plasma frequency and the width
of the Drude peak, whereas $\omega_j$, $\gamma_j$, and $S^2_j$ are
the center-peak frequency, the width, and the mode strength for
the $j$-th Lorentz harmonic oscillator (h.o.), respectively.
$\sigma_1(\omega)$ is then obtained from $\sigma_1(\omega)=\omega
\epsilon_2(\omega)/4\pi$.

The fit of all spectra were extended up to 40000 cm$^{-1}$.
Besides the Drude contribution, five Lorentz h.o.'s for each
compound are required to fit the finite-frequency features. This
is explicitly shown in the inset of Fig.~\ref{Sig1} for DyTe$_3$,
where the single fit components are displayed. The three low
frequency oscillators allow to reproduce the rather broad
absorption, ascribed to the SP peak. This choice is motivated by
the fact that, for all samples, the SP peak cannot be fitted with
a single Lorentzian oscillator. There is indeed the presence of
low- and high- frequency shoulders, each described by a Lorentz
h.o., which overlap to a background defined by a broad h.o. In
LaTe$_3$ and CeTe$_3$ the shoulder at the low frequency side of
the SP peak almost merges with the high frequency tail of the
Drude component. Therefore, the SP-peak in each compound can be
thought as composed by the superposition of several excitations,
which mimic the continuous distribution of gap values, as observed
by ARPES \cite{ARPES2,ARPES3,ARPES1,ARPES4, XPSgap}. The remaining
two high frequency h.o.'s account for the optical (electronic
interband) transitions.

\begin{figure}[!tb]
\center
\includegraphics[width=8.5cm]{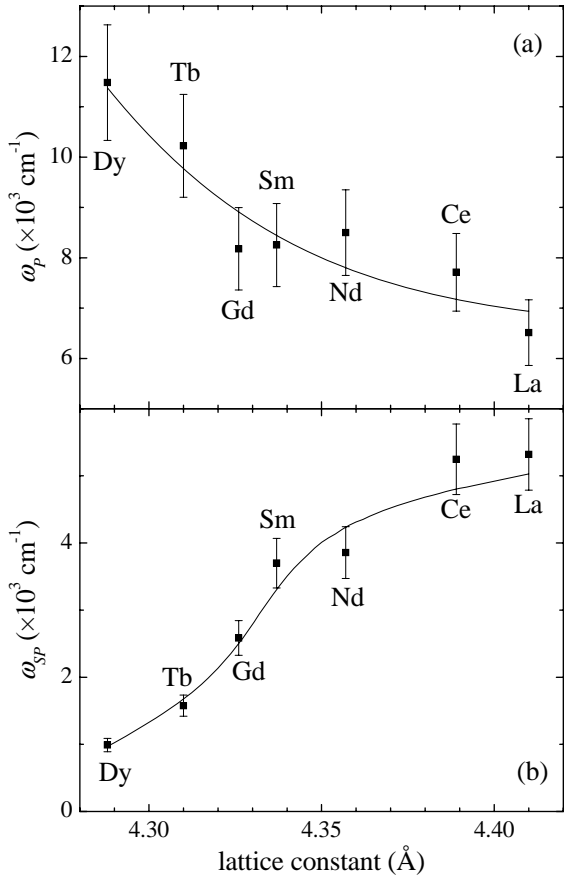}
\caption{Plasma frequency (a) and single-particle peak frequency
(b) as a function of the in-plane lattice constant
\cite{LattConst}. The rare-earth atom for each compound is shown
in the plot. The solid lines are guides to eyes.} \label{DvsSP}
\end{figure}

There are several interesting parameters, which can be extracted
from the fit. First of all, the plasma frequency $\omega_P$, the
square of which represents the total spectral weight of the Drude
peak. The larger $\omega_P$ is, the higher is the metallic degree
of the system. It is also worth to recall that $\omega_P^2 \propto
n/m^{\ast}$, where $n$ is the charge-carriers' density and
$m^{\ast}$ is the carriers' effective mass. The obtained values of
$\omega_P$ are plotted in Fig.~\ref{DvsSP}a, as a function of the
lattice constant $a$ \cite{LattConst}.
An increase of $\omega_P$ going from LaTe$_3$ to DyTe$_3$ is well
evident.

As to the SP peak, since it is composed by three Lorentz h.o.'s, we
define the averaged quantity $\omega_{SP}$:
\begin{equation}
\omega_{SP}=\frac{\sum_{j=1}^3 \omega_j S_j^2}{\sum_{j=1}^3
S_j^2},
\end{equation}
which represents the center of mass of the SP excitation and thus
provides an optical estimate for the CDW gap. In
Fig.~\ref{DvsSP}b, $\omega_{SP}$ is plotted as a function of $a$
\cite{LattConst}. The decrease of $\omega_{SP}$ (thus of the
average gap value) is well evident on going from LaTe$_3$ to
DyTe$_3$, oppositely to the observed increase of $\omega_P$ (thus
of the Drude spectral weight, Fig.~\ref{DvsSP}a). Another
interesting quantity associated to the SP-peak is the total
spectral weight $S_{SP}^2$ of the (broad) SP excitation, which is
defined as:
\begin{equation}
S_{SP}^2=\sum_{j=1}^3 S_j^2.
\end{equation}

At this point we are then equipped for a thorough discussion of
our findings. Before addressing in detail the evolution of the
electronic properties across the rare-earth series, we first focus
our attention on the relationship between the CDW and the normal
state (NS) properties for each compound. Although the
high-temperature NS cannot be studied experimentally since
$T_{CDW}$ is anticipated to be above the melting point of the
material, we can nevertheless draw some predictions by exploiting
our data and the band-structure calculations \cite{Band,Band2}.
The Te $p_x$ and $p_y$ orbitals have a strong 2D character (i.e.,
they have a negligible dispersion along the direction orthogonal
to the Te layers). These bands are well approximated by a
tight-binding model \cite{ARPES2} in which, taking into account
the stoichiometry, the Fermi level lies above half-filling (i.e.,
at 3.25 eV below the top of the bands) \cite{SpecHeat}, indicating
that charge carriers are hole-like. Considering a parabolic
expansion of the 2D bands around their maximum, the Fermi energy
$E_F=3.25$ eV, within a 2D free-holes model, is then related to
the effective mass $m_{NS}$ by:
\begin{equation}
E_F=\frac{\pi \hbar^2 n_{2D}}{m_{NS}} \label{mNS}
\end{equation}
where $n_{2D}$ is the 2D hole density in NS. This latter quantity
is determined by assuming 1.5 holes for each tellurium atom within
the Te layers and 2 Te atoms for each square unit within the 2D
layers (i.e., $n_{2D}=3/a^2$). From $n_{2D}$ and $E_F$, we achieve
$m_{NS}$ for each sample by exploiting eq. (\ref{mNS}). The
$m_{NS}$ values are reported in Table \ref{Tab}.
\begin{table}[!t]
\centering
\begin{tabular*}{\columnwidth}{@{\extracolsep{\fill}}cccccc}
\\\hline\hline
Sample & $m_{NS}$ & $n_{NS}$ & $m_{CDW}$
& $n_{CDW}$ & $\Phi=\omega_P^2/S_{NS}^2$ \\
\hline
LaTe$_3$ & 1.1(1) & 12(2) & 0.56(6) & 0.13(1) & 0.021(4)\\
%\hline
CeTe$_3$ & 1.1(1) & 15(3) & 0.56(6) & 0.18(2) & 0.024(5)\\
%\hline
NdTe$_3$ & 1.2(1) & 14(3) & 0.59(6) & 0.23(2) & 0.032(6)\\
%\hline
SmTe$_3$ & 1.2(1) & 17(4) & 0.59(6) & 0.22(2) & 0.025(5)\\
%\hline
GdTe$_3$ & 1.2(1) & 13(3) & 0.59(6) & 0.22(2) & 0.032(6)\\
%\hline
TbTe$_3$ & 1.2(1) & 17(4) & 0.60(6) & 0.34(3) & 0.041(8)\\
%\hline
DyTe$_3$ & 1.2(1) & 13(3) & 0.60(6) & 0.43(4) & 0.065(13)\\
\hline\hline
\end{tabular*}
\caption{Effective mass $m_{NS}$ and carriers' density $n_{NS}$ in
the normal state and effective mass $m_{CDW}$ and carriers'
density $n_{CDW}$ in the CDW state for all samples. Carriers'
densities and effective masses are given in carriers/cell and
$m_e$ units, respectively. The last column reports the ratio
$\Phi$ between the Drude spectral weights (see text) in the CDW
and normal state. Experimental uncertainties in the last digits
are given in brackets.} \label{Tab}
\end{table}

The $m_{NS}$ values allow us to check our data in a
self-consistent manner. If we assume the conservation of the
spectral weight between the CDW and the normal state, and that
there will be no SP peak in the (hypothetical) NS, the Drude
contribution in NS would then have a total spectral weight given
by $S^2_{NS}=\omega^2_P+S^2_{SP}$ (Fig. 3). Therefore, from
$S_{NS}^2\sim n_{NS}/m_{NS}$ we can estimate the three-dimensional
carriers density $n_{NS}$ in NS, as reported in Table \ref{Tab}
for each compound. This quantity can be compared with the
carriers' density $n_{ch}$ obtained from the chemical counting.
Since the 3D unit cell contains four Te layers (i.e., 8 Te atoms)
with 1.5 holes for each Te atom, we get $n_{ch}=12$ holes/cell.
The fair agreement (at least within the experimental
uncertainties) between $n_{NS}$ and $n_{ch}$ is well evident in
Table \ref{Tab}. This finding makes us confident about the
reliability of our analysis.

We now consider the free charge carriers surviving in the CDW
state. We can estimate their effective mass $m_{CDW}$ from the
existing specific heat data on LaTe$_3$. The Sommerfeld
$\gamma$-value of the linear term in the specific heat is
$\gamma=0.0011$ J mol$^{-1}$ K$^{-2}$ for LaTe$_3$
\cite{Ru,SpecHeat}. We assume a 2D free-holes scenario for the
ungapped carriers also in the CDW phase. Within this approach,
$\gamma$ and $m_{CDW}$ are related by:
\begin{equation}
\gamma=\frac{\pi k_B^2 m_{CDW} a^2}{3 \hbar^2}. \label{mCDW}
\end{equation}
Since $\gamma$ is difficult to extract from heat capacity
measurements for the magnetic members of the rare earth series, we
use the value of LaTe$_3$ for all samples. Bearing in mind the
small chemical and structural changes occurring across the
rare-earth series, this assumption appears reasonable. Therefore,
we can estimate $m_{CDW}$ from eq. (\ref{mCDW}) for all samples,
as reported in Table \ref{Tab}. It turns out that the
free-carriers' mass decreases by a factor 2 on entering the CDW
state. This finding can be explained by noting again that in
$R$Te$_3$ the relevant electronic states correspond to the two
orthogonal bands deriving from the Te $p_x$ and $p_y$ orbitals.
The band structure, while nearly 2D, is anisotropic in the plane.
Therefore, $m_{NS}$ and $m_{CDW}$ should be considered as
parameters describing the average curvature of the free charge
carriers bands in the normal and CDW state, respectively. In this
respect, our estimates of the effective masses represent a
harmonic average of the mass per carrier, i.e.,
$1/m_{NS}=1/n_{NS}\sum_{i}1/m_{i}$ and
$1/m_{CDW}=1/n_{CDW}\sum_{i}1/m_{i}$, where the two sums run over
the ungapped states in the normal and CDW phases, respectively.
The difference in the $m_{NS}$ and $m_{CDW}$ values thus indicates
that in the CDW phase the average band curvature is larger than in
NS. Consequently, the CDW mainly gaps ``heavy'' carriers (i.e.
those belonging to states with small band-curvature). The
remaining ungapped carriers thus occupy states with a larger
band-curvature, resulting in smaller effective mass. Since states
with small band-curvature are flatter (i.e. they have a smaller
dispersion) than those with large band-curvature, one can expect
the former to give a larger contribution to the density of states
than the latter. Therefore, it is not at all surprising that the
CDW tends to gap states with the small band-curvature. This way
there are more states which can be gapped and the CDW energy-gain
is larger.

The knowledge of $m_{CDW}$ and $\omega_P$ allows us to determine
the density $n_{CDW}$ of the ungapped carriers in the CDW phase.
The $n_{CDW}$ values for each sample are reported in Table
\ref{Tab}. The large difference between $n_{NS}$ and $n_{CDW}$ and
consequently the dramatic reduction of the number of ungapped
charge carriers indicate the strong effect of the CDW formation; a
large part of the charge carriers takes part in the formation of
the CDW condensate. That a significant part of the FS is gapped in
the CDW phase, is confirmed by ARPES data from which the fraction
of the ungapped FS can be estimated to be 10-20\%
\cite{ARPES1,ARPES2,ARPES3,ARPES4}.

In this respect, the quantity
$\Phi=\omega_P^2/(\omega_P^2+S_{SP}^2)$ is of interest, since it
represents the ratio between the spectral weight of the Drude peak
and the total spectral weight of the Drude term and the SP peak
(Table I). $\Phi$ roughly measures, as previously shown in Ref.
\onlinecite{Perucchi1}, the fraction of the ungapped FS area
(i.e., those parts of FS which are not affected by the CDW state).
Since the effective mass of the free carriers is different in CDW
and NS, $\Phi$ does not simply reduce to $n_{CDW}/n_{NS}$ and it
is larger than the latter quantity by a factor given by
$m_{NS}/m_{CDW} \sim 2$. Although the $\Phi$ values seem to
significantly underestimate the fraction of the ungapped FS
compared to the estimate evinced from ARPES data
\cite{ARPES1,ARPES2,ARPES3,ARPES4}, they at least qualitatively
confirm that a large portion of FS is gapped. The difference
between ARPES and optics, as far as the gapping of FS is
concerned, may also be reconciled in part by taking into account
the energy resolution of the ARPES data \cite{noteARPES}.
\begin{figure}[!t]
\centering
\includegraphics[width=8.5cm]{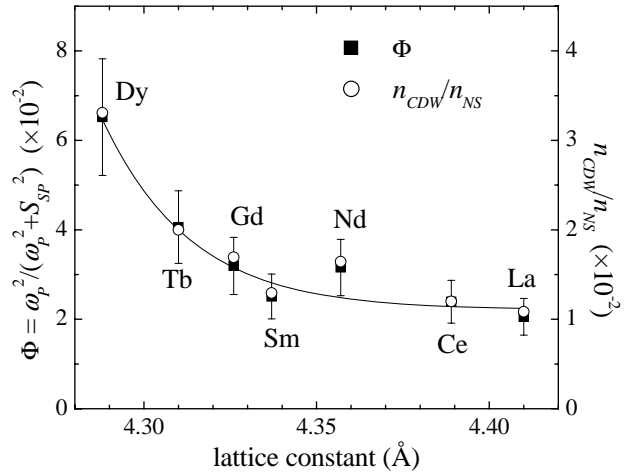}
\caption{Ratios $\Phi=\omega_P^2/(\omega_P^2+S_{SP}^2)$
(indicating the fraction of the ungapped FS) and $n_{CDW}/n_
{CDW}$ as a function of the in-plane lattice constant
\cite{LattConst}. The left and right scales differ by a factor of
2. The rare-earth atom for each compound is shown in the plot. The
solid line is a guide to eyes.} \label{Phi}
\end{figure}
We now turn to the evolution of the CDW state across the
rare-earth series. First of all, the reduction of $\omega_{SP}$ on
decreasing $a$ (Fig. 3b) may be considered as an indication for
the diminishing impact of the CDW state, when going from La to Dy.
Such a gap reduction is also consistent with the ARPES data
\cite{XPSgap}, showing a gap of about 400 meV in CeTe$_3$
\cite{ARPES2,ARPES3} and of 200 meV in SmTe$_3$ \cite{ARPES1}. In
recent ARPES experiments \cite{ARPES4} a reduction of the maximum
gap-value with decreasing $a$ was also observed for several
compounds of the $R$Te$_3$ series. It is important to remark that,
differently from ARPES where the maximum gap-value can be
determined, our optical estimate provides a sort of averaged value
for the CDW gap over the whole FS.

In Fig.~\ref{Phi}, $\Phi$ (Table I) is plotted as a function of
the lattice constant $a$ \cite{LattConst}. The portion of the
ungapped FS increases on decreasing $a$ and has a more gradual and
less scattered behavior than $\omega_P$ (Fig. 3a). Since $\Phi$
corresponds to the ratio between spectral weights, it is less
affected by possible uncertainties of the $\sigma_1(\omega)$
absolute value. In the same figure, the ratio $n_{CDW}/n_{NS}$
between the carrier's densities in the CDW and NS (see Table I) is
also plotted as a function of $a$. As discussed above, the two
quantities have the same dependence on $a$ and just differ in
absolute value by a factor $\sim2$ (Fig. 4), coming from the
effective mass ratio. The increase of $\Phi$ with decreasing $a$
is quite abrupt for rare earths beyond the Sm-containing compound,
indicating the onset of a chemical pressure effect for $a \leq
4.34$ \AA. On the other hand, the effective mass is almost
constant across the rare-earth series so that the band-shape and
the inertia of the free charge carriers are only subtly affected
by the lattice compression. Therefore, this seems to rule out a
large change in the band width. Nevertheless, one cannot exclude a
priori that a yet subtle narrowing of the bands with increasing
lattice constant could lead to a better nesting and to differences
in the amount of the nested FS.

In this context, it is then quite natural to assume that the
reduction of $\omega_{SP}$ (Fig. 3b) on going from LaTe$_3$ to
DyTe$_3$ could be ascribed to a suppression of the nesting
condition, due to the changes in FS, because of the lattice
compression. In this scenario, the increase of the ungapped
portions of FS with decreasing $a$ (Fig. 4) leads to an enhanced
optical contribution due to the free charge carries. The strong
decrease of $\omega_{SP}$ is consistent with the reduction of the
perfectly-nested regions (where the CDW gap is close to its
maximum value) in favor of the non-perfectly-nested regions (where
the gap is close to zero).

\begin{figure}[!t]
\centering
\includegraphics[width=8.5cm]{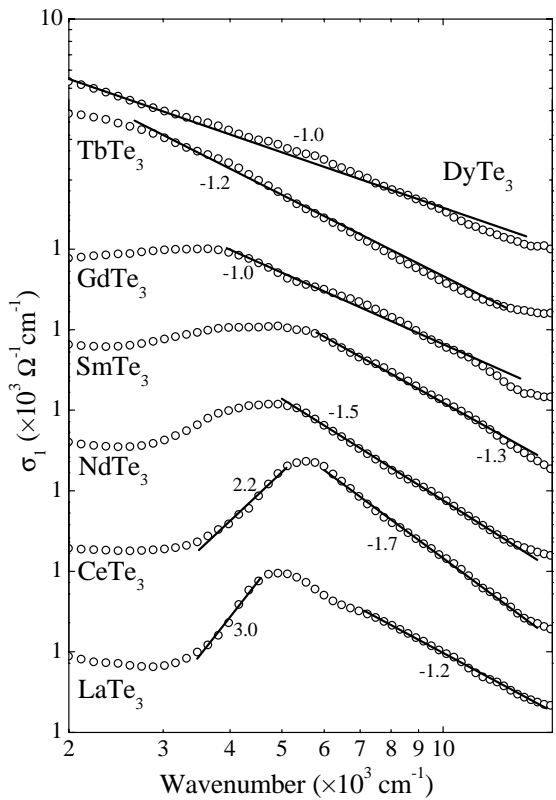}
\caption{$\sigma_1(\omega)$ of $R$Te$_3$ ($R$=La, Ce, Nd, Sm, Gd,
Tb and Dy) plotted on a bi-logarithmic scale. The y axis
logarithmic scale is vertically shifted for the sake of clarity.
The scale is the same for all samples and the scale-offset is
shown for each spectrum. The solid lines are power-law fits to the
data (the exponents are given in the figure).} \label{Power}
\end{figure}
Finally, a power-law ($\sigma_1(\omega)\sim\omega^{\eta}$)
behavior for the high frequency tail of the SP peak is observed in
all samples. This is shown in Fig.~\ref{Power}, where
$\sigma_1(\omega)$ is plotted on bi-logarithmic scale. The
power-law behavior extends over a rather limited energy interval,
which at most is one order of magnitude wide for the Dy compound.
Power law scaling is indicative of quasi 1D behavior
\cite{vescoliscience,schwartz_electrodynamics}. The observed
exponents, although not precisely determined, give indications as
to the mechanism behind CDW formation in these materials. In the
following analysis we make the case that electron-electron
interactions play a crucial role in the CDW formation in
$R$Te$_3$.

The traditionally invoked mechanism for CDW formation is the
electron-phonon coupling. In that case there are three energy
scales to be considered for optical absorption: the typical phonon
frequency $\omega_0$, the single particle (or Peierls) gap
$\Delta$, and the frequency $\omega$ at which the measurement is
done. As anticipated above, the advocated FS of the material
\cite{ARPES2} consists of two sheets of open FS's of a quasi 1D
material (associated to the $p_x$ and $p_y$ orbitals,
respectively). The measured vector for the CDW modulation is very
close to the vector that corresponds to the nesting of the two
sides of this quasi 1D FS. Attributing the one dimensionality to
the fact that $R$Te$_3$ have a nearly perfect nested quasi-1D FS
is of particular relevance here, since the charge transfer
integral ($t_{perp}$) along the transverse direction (i.e.,
describing the hopping between the $p_x$ and $p_y$ orbitals) is
not small and is much larger than the temperature of the
measurements. Indeed, $t_{perp}>T$ would normally lead to coherent
transverse hopping, so that FS would have significant warping in
the transverse direction and the material would not be 1D anymore.
The warping of FS would loose its relevance only at
$\omega>t_{perp}$. However, this is not the appropriate situation
for $R$Te$_3$, since $t_{perp}= 0.37$ eV \cite{ARPES2,Band}, while the power law behavior is observed for frequencies $\omega>0.2$ eV. But
if nesting is strong and occurs with a well defined $\vec{q}$
vector, then the system still acts as a 1D system would
essentially do. The 1D character, indicated by the high frequency
power law behavior of $\sigma_1(\omega)$ (Fig. 5), may then
persist even for $\omega\leqslant t_{perp}$, provided that one
looks at phenomena involving the nesting wave-vector.

In the case of a 1D material, one would get different exponents
for the optical behavior above the Peierls gap depending on the
hierarchy of the energy scales, pointed out above. If one is in
the so-called adiabatic limit $\omega_0 \ll \Delta$ where the
phonon frequency is quite small, then one can assimilate the
potential created by the phonons to either a periodic (with
wave-vector $\vec{q}$) static deformation, if $\omega \ll
\omega_0$, or to a quenched disorder varying in space, if $\omega
\gg \omega_0$. In the first case the conductivity can easily be
computed by looking at the scattering on the static periodic
potential with a wave-vector $\vec{q}$, the CDW modulation vector,
using the techniques explained in Ref.
\onlinecite{giamarchi_book_1d}. One finds that $\sigma_1(\omega)
\propto \omega^{K_\rho - 4}$ where $K_\rho < 1$ is the Luttinger
liquid parameter. $K_\rho = 1$ for a Fermi liquid or if
interactions are weak and decreases with increasingly repulsive
interactions. It is clear that such a result would not be
compatible with the data, making it hard to have the standard
adiabatic mechanism for the CDW formation. The other case, when
the phonon potential is viewed by the electron as a quenched
disorder, is also not quite compatible with the data. Indeed in
that case one finds $\sigma_1(\omega) \propto \omega^{K_\rho - 3}$
\cite{giamarchi_book_1d}. A weakly interacting system would thus
give exponents of order $-2$ or slightly below, in somewhat closer
but yet very unsatisfactory agreement with the data. It thus seems
unlikely that the data are explained by the standard scattering
over a distortion due to adiabatic phonons. The data would on the
contrary be in reasonable agreement with the so-called
antiadiabatic limit $(\Delta,\omega) \ll \omega_0$, in which one
can integrate over the phonon field. The phonon fluctuations
introduce then an effective interaction between the particles.
This interaction corresponds to an umklapp scattering, which gives
\cite{Lutt,giamarchi_book_1d}:
\begin{equation} \label{eq:umk}
 \sigma_1(\omega) \propto \omega^{4K_\rho - 5}.
\end{equation}
This would lead to exponents $\eta$ close or slightly smaller than
$-1$, in much better agreement with the data. It is however
extremely unlikely to be able to find phonons of such high
frequency, since one would need $\omega_0 > 1$ eV. Consequently,
phonons alone could not explain the observed optical data and
power laws.

A much more probable source for such an umklapp scattering leading
to eq. (\ref{eq:umk}) is the direct interaction between electrons.
Such an interaction can directly produce an umklapp scattering,
allowing the transfer of two particles from one sheet of the FS to
the other and thus transferring $4\vec{q}$ to the lattice
\cite{giamarchi_book_1d}. Note that the advocated wave-vector
\cite{ARPES2} for the CDW modulation is indeed giving a value for
$4\vec{q}$ close to a reciprocal lattice vector, and thus allowing
such umklapp processes to be effective. An umklapp process leading
to a power law behavior in $\sigma_1(\omega)$ and in other
response functions was theoretically predicted for strict 1D
systems within the Tomonaga-Luttinger-liquid scenario \cite{Lutt}
and observed experimentally in the 1D Bechgaard salts with
$\eta=-1.3$
\cite{schwartz_electrodynamics,vescoliscience,LeoEPJB}.
Furthermore, we identify in $\sigma_1(\omega)$ of the Ce and La
compounds a low-frequency tail of the SP peak following a
power-law behavior as well (Fig. 5). The resulting exponent ranges
between 2.2 and 3, which again compares quite well with the value
of 3 predicted for a 1D Mott insulator for which umklapp is the
dominant source of scattering \cite{LeoEPJB}.

Even though the exponents for the high-frequency power-law
observed in our data do not show a defined trend across the
rare-earth series, their values range between -1.7 and -1.0. This
is in decent agreement with such a scenario where interactions and
umklapp would be responsible for the observed behavior. This
strongly suggests that interactions rather than a standard
electron-phonon mechanism could play, in the rare-earth
tri-tellurides, a crucial role in the CDW formation as well. Of
course more studies both theoretically and experimentally would be
useful to ascertain the respective roles of the interactions and
of electron-phonon coupling with respect to the CDW formation.
Theoretically, a more careful treatment of the effects of the
transverse warping of FS would be clearly needed, not only for the
high energy behavior but even more for the low frequency part of
the optical conductivity, much below the single particle peak.

\section{Conclusions}

In summary, we reported on the first optical measurements of seven
different rare-earth tri-tellurides. Our data allow for a detailed
analysis of both the Drude contribution, ascribed to the free
charge carriers resulting from the presence of ungapped regions of
FS, and the SP peak, due to carriers' excitation across the CDW
gap. On decreasing the lattice constant, a slight enhancement of
the metallic contribution and a simultaneous reduction of the CDW
gap are observed. We propose that this effect might be due to a
quenching of the nesting condition driven by a modification of FS
because of the lattice compression. We also observe power-law
behaviors in $\sigma_1(\omega)$, typical of a
Tomonaga-Luttinger-liquid system, which emphasize a non negligible
contribution of 1D correlation effects in the physics of these 2D
compounds. This also anticipates that interactions and umklapp
processes could play a significant role in the CDW formation in
these compounds.

\begin{acknowledgments}
The authors wish to thank J. M\"uller for technical help, and V. Brouet and M.
Lavagnini for fruitful discussions. This work has been supported
by the Swiss National Foundation for the Scientific Research
within the NCCR MaNEP pool. This work is also supported by the
Department of Energy, Office of Basic Energy Sciences under
contract DE-AC02-76SF00515.
\end{acknowledgments}


\begin{thebibliography}{99}
\bibitem{footnote} The $Cmcm$ space group is orthorhombic. However, the in-plane
lattice constants $a$ and $c$ differ by approximately 0.1\% (in the standard
space group setting the $b$-axis is perpendicular to the Te planes). For simplicity,
in our subsequent analysis we treat the material as essentially tetragonal,
characterized by an in-plane lattice constant $a$.
\bibitem{XPSgap} The gap values determined from the ARPES experiments are always
referred with respect to the Fermi energy. Assuming the Fermi energy in the
middle of the gap, means that the optically determined gap is about twice as
large as the ARPES one.
\bibitem{noteARPES} Fermi surface, reconstructed from ARPES data, corresponds
to integrated signals up to an energy window around the Fermi energy.
Therefore, the gapping of the Fermi surface will only extend over regions
corresponding to gap values larger than the experimental resolution, while
the real gap extends further around the Fermi surface than its ARPES mapping
appears to show.
\end{thebibliography}
\end{document}